# Tunable thermal conductivity of ferroelectric P(VDF-TrFE) nanofibers via molecular bond modulation

Lan Dong(董岚),[1,2] Bohai Liu(刘博海),[3] Yuanyuan Wang(王元元),[1,2†] Xiangfan Xu(徐象繁)[3†]

[1]School of Energy and Materials, Shanghai Polytechnic University, Shanghai 201209, China

[2]Shanghai Engineering Research Center of Advanced Thermal Functional Materials, Shanghai Polytechnic University, Shanghai 201209, China

[3]Center for Phononics and Thermal Energy Science, China-EU Joint Center for Nanophononics, School of Physics Science and Engineering, Tongji University, Shanghai 200092, China

[†]Correspondence and requests for materials should be addressed to Y.W. (email: wangyuanyuan@sspu.edu.cn ) or to X.X. (email: xuxiangfan@tongji.edu.cn)

**ABSTRACT:** The dipoles in ferroelectric copolymer P(VDF-TrFE) can be driven by electric field, introducing phonon transport modulations through polarizing molecular chains. The thermal conductivity in single 75/25 P(VDF-TrFE) nanofibers is found to increase with electric field related phonon renormalization, resulted from change in vibrational assignment excited by polarization process. This is evidenced by a direct change of bond energy and bond length in 75/25 P(VDF-TrFE) nanofibers from Raman characterization under polarization electric field. The experimental results provide further intuitive evidences that the size of ferroelectric polymers could directly affect the ferroelectricity from the size-dependent thermal transport measurement.

The spontaneous electrical polarization of ferroelectrics is crucial to the semiconductor mechanism research and many of current applications which relates to the tunable Field-Effect Transistor[1, 2], the non-volatile random access memories[3-5], etc. Ferroelectrics usually have piezoelectric properties that generated tremendous interests in electrical field applications including energy storage, actuators and sensors[6-9]. Rochelle salts[10] was discovered in the last century, whose polarization undergoes a hysteresis loop with electric field[11-13], similar to the magnetic hysteresis loop in ferromagnetic materials[14, 15]. Rochelle salts founded a new stage among numerous ferroelectric researches and also developed the field which has been dominated by inorganic ferroelectrics like Pb(Zr$_x$Ti$_{1-x}$)O$_3$ (PZT)[16, 17], BaTiO$_3$[18], etc. Inorganic ferroelectrics have been exploited to develop the energy-efficient technology for micro-nano logic device[19] as well as non-volatile random storage. However essentially poisonous and preparation obstacles make the applications of inorganic ferroelectrics severely restricted.

As polymer ferroelectrics, polyvinylidene fluoride (PVDF) and its copolymer poly(vinylidenefluoride-co-trifluoroethylene) (P(VDF-TrFE)) are both fluorine-based polymers which are widely used in modern electronics and electrical systems due to their high dielectric constant and breakdown strength. Recent years, many researches focused on improving the energy density of the PVDF and its copolymer which could further optimize the breakdown strength and discharge efficiency in capacitive energy storage[20, 21]. One key research topic is the physical properties and stability of polyvinylidene fluoride and its copolymer under extreme condition like ultra-high voltage with voltage gradient approaching breakdown threshold. In addition, heat dissipation is a critical issue that should to be addressed immediately to improve long-term reliability and performances of electronics. PVDF and its copolymers need to pay attention to their heat dissipation properties while optimizing their electric performance. Therefore, the research on the thermal transport of PVDF and its copolymer P(VDF-TrFE) under high voltage becomes particularly important[22]. The regulation of polymer thermal transport is affected by micro/ nano transport mechanisms, including hydrogen bond[23], phonon coupling[24], interface thermal transport[25], etc. Therefore, the regulation mechanism of thermal transport in ferroelectric polymers needs to be more detailed.

Semi-crystalline ferroelectric PVDF and P(VDF-TrFE) hold spontaneous polarization based on its intrinsic structural polarization on the both side of the C-C backbone[26, 27]. In principle, there are four crystallize phases of fluorine-based ferroelectric polymers with different chain conformations, e.g., $\alpha$-phase, $\beta$-phase, $\gamma$-phase and $\delta$-phase[28-33]. The $\alpha$-phase, the dominate phase in PVDF conformation, is non-polar with very weak ferroelectricity. Compared to weak ferroelectricity in $\alpha$-phase, $\beta$-phase has an all-trans planar-zigzag chain conformation, which is highly polar structure. Among the four crystalline phases, $\alpha$-phase is the non-polar phase but the most stable phase, therefore it is the most likely polymerization during the synthesis process of fluorine-based ferroelectric polymers. While for the $\beta$-phase (polar phase), the requirements for polymerization of fluorine-based polymers are more stringent and complex because of the instability of $\beta$-phase. P(VDF-TrFE) is a typical PVDF copolymer with excellent ferroelectric properties, with its chain conformation mainly composed of β-phase. TrFE is the key point for the formation of β-phase, the higher mole fraction of TrFE can achieve the better ferroelectricity of P(VDF-TrFE)[34].

It is the inherent property of ferroelectric system that the dipole can be driven by an applied polarization electric field. The polymer ferroelectrics have intrinsic dipoles whose polarization direction is perpendicular to the C-C backbones. The electrostatic force generated by the applied electric field is believed to affect the bond length, bond angle and the atomic vibration modes in ferroelectric system. However, the change of chain configuration in not easy to be observed in bulk ferroelectric polymers due to the complicated and confused molecular chain structures of polymers. In order to explore the coupling relationship between chain configuration and applied polarization electric field, P(VDF-TrFE) nanofibers are prepared in-situ on Micro-Electro-Mechanical System (MEMS) devices using electrospinning process, suitable for thermal transport measurement. The electrospining process forces the molecular chain align along the longitudinal direction of nanofibers. Such structural optimization is helpful for further studying the change of the bond length, bond angle and the atomic vibration modes in ferroelectric polymers, which is believed to affect the thermal properties of the ferroelectric nanofiber but however not studied experimental yet.

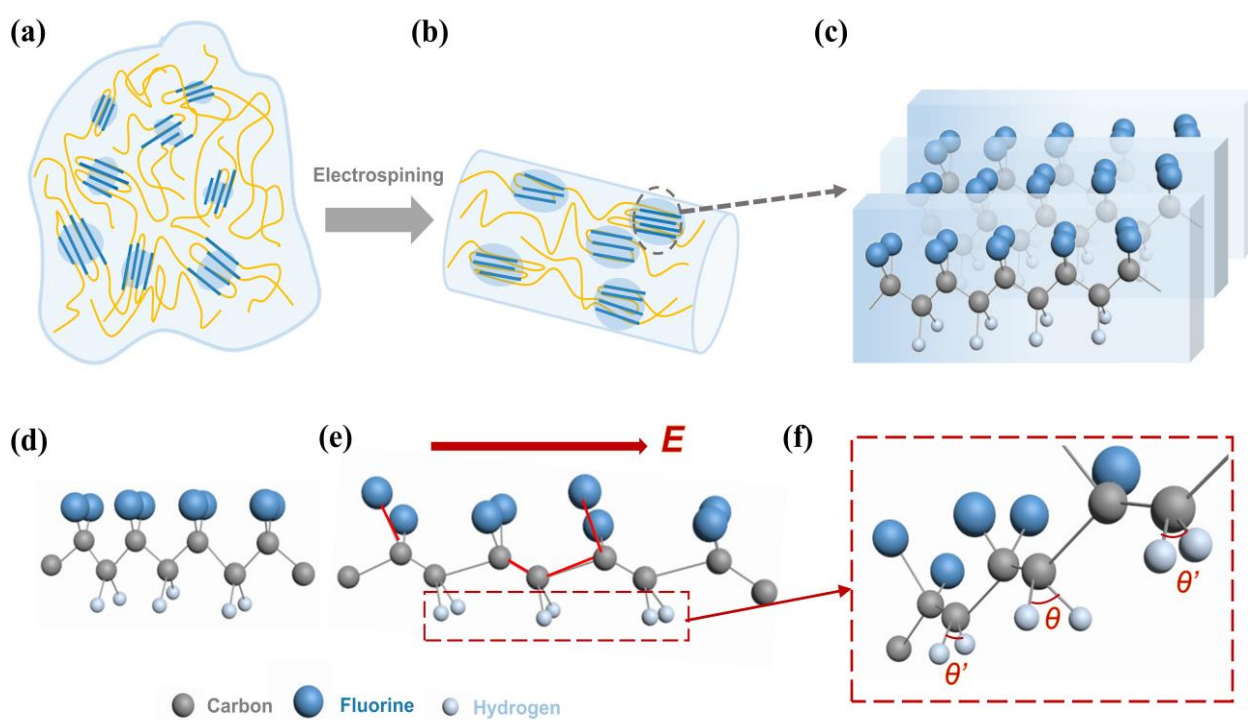

**Figure1** Schematic of molecular conformation under polarization electric field and the morphology details of ferroelectric P(VDF-TrFE). (a) Isotropic bulk P(VDF-TrFE) system with the chain ends, entanglements, voids and defects; (b) Schematic of molecular chain orientation of P(VDF-TrFE) nanofiber after electrospinning process with no polarization electric field applied. The light blue circle areas represent the crystalline regions, the yellow lines represent the amorphous regions; (c) The arrangement of β-phase in crystalline regions of P(VDF-TrFE) nanofiber; (d) The molecular chains of typical β-phase of P(VDF-TrFE) under zero field; (e) The change of bond energy, bond length and bond angle of β-phase chains under polarization electric field (red arrow), the red lines represent the change of bond length of C-C bond and $CF_2$ bond, the red dashed rectangle represents the change of angles of $CH_2$ bonds. (f) Enlarged image exhibits the changed angles of $CH_2$ bonds.

The bulk semi-crystalline polymers contain both aligned crystalline and chain-oriented amorphous phases, with the crystalline region surrounded by the random amorphous regions[35]. Bulk P(VDF-TrFE) contains the chain ends, entanglements, voids and defects that inhibit the inter and intra molecular thermal transport (shows in Fig. 1(a)), resulting in inferior thermal conductivity (0.1-0.3 $Wm^{-1}K^{-1}$) comparing to that in aligned molecular chains and nanofibers[36, 37]. Electrospining is a mature and widely used technology for polymer nanofiber preparation, with its electric voltage generally ranging from thousands of volts to tens of thousands of volts. Under the electrostatic force created by the high voltage, the polymer solution is first stretched into an ellipsoid, and then nanofibers with diameters in the order of micrometers or nanometers are prepared at the front end of the ellipsoid. The polymer nanofiber prepared by electrospining technique could help align the molecular chains (anisotropic nanofibers) and decrease the phonon scattering which is the main factor restricting the thermal conductivity of polymers (shows in Fig. 1(b)). Crystalline phase in semi-crystalline polymers appears as the folded chain regions. For copolymer P(VDF-TrFE), it has well-arranged H-F dipoles, whose polarization direction is perpendicular to the C-C backbone (details could find in Fig. 1(c)). In this experiment, single P(VDF-TrFE) nanofibers (the molar ration of P(VDF-TrFE) is 75/25, which was purchased from PIEZOTECH company) are prepared across the two thermometers of MEMS device. Considering the anisotropy of molecular chain orientation caused by electrospinning process, an applied polarization electric field ($E$) was provided parallel to the orientation direction of the C-C backbones, which is also the axial direction of the P(VDF-TrFE) nanofibers. Fig. 1(d) exhibits the β-phase molecular chains of P(VDF-TrFE) without polarization electric field ($E = 0$), whose dipole direction is perpendicular to the arrangement direction of the C-C chain. The bond length of C-C band and side chains will change to some extent (represent by the red lines in Fig. 1(e)) under polarization electric field. The variation of bond length is consistent with assumptions in the previous work[38].

The changes of bond energy, bond length and bond angle are the direct factors affecting the thermal transport of nanostructures, especially in polymer nanofibers. In order to observe the change of thermal conductivity caused by the polarization electric field mentioned above, thermal bridge method was used to characterize the axial thermal conductivity of P(VDF-TrFE) nanofibers with 75/25 molar ratio. The whole suspended MEMS device was placed in a cryostat with high vacuum on the order of $1\times10^{-5}$ Pa to reduce the effect of thermal convection. Considering the change in thermal conductivity of a single P(VDF-TrFE) nanofiber would be much lower than the measurement sensitivity of traditional thermal bridge method[39-43], we adopted the differential circuit configuration due to its advanced measurement sensitivity approaching to ~10 pW/K[44].

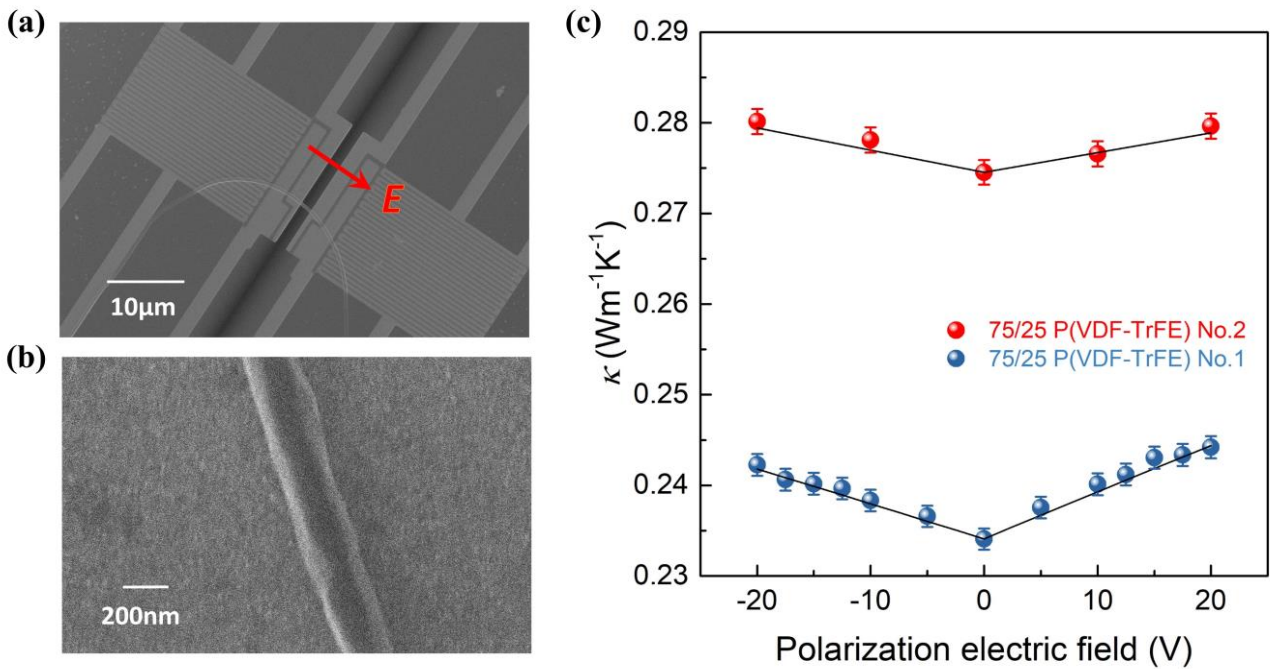

**Figure 2** Polarization tunable thermal transport of P(VDF-TrFE) nanofibers with 75/25 molar ratio at $T$ = 300 K. (a) SEM image of a single 75/25 P(VDF-TrFE) nanofiber which suspended on the MEMS device for thermal transport measurement, the scale bar is 10μm. The middle electrodes of MEMS device are used to add the polarization electric field. The polarization electric field is provided by the Keithley 6430 and the red arrow indicates its direction. (b) Enlarged SEM image of the same P(VDF-TrFE) nanofibers in Fig. 2(a). The scale bar is 200nm. (c) Thermal conductivities of the No.1 and No.2 nanofibers with the change of polarization electric field. As the positive electric field approach to 20 V, the thermal conductivity of No.1 sample increases by 4.3% when compared with the unpolarized thermal conductivity, while for negative electric field (up to -20V) the increase of thermal conductivity reaches 3.5%. Thermal conductivity increases by 1.9% with the highest positive electric field (+20 V) and 2.0% with the highest negative electric field (-20 V) for the No.2 sample. The black solid lines are fitted by formula (1).

Figure 2(a) is the SEM image of a suspended single P(VDF-TrFE) nanofiber with 75/25 molar ratio (Sample: 75/25 No.1). The nanofiber was fabricated across the two Pt/SiN$_x$ thermometers by electrospinning technology[45]. After electrospinning the P(VDF-TrFE) nanofiber need to be annealed at 140°C in high purity nitrogen environment to improve its crystallization of β-phase [46]. The red arrow illustrates the direction of the polarization electric field. It is the key step where the tunable molecular conformation gradually affects the phonons transport along the axis of the P(VDF-TrFE) nanofiber. To avoid the interaction between pluralities of nanofibers, we prepared single suspended nanofiber that is extremely difficult for electrospinning to measure the polarization tunable thermal conductivity. Fig. 2(b) presents the enlarged nanofiber morphology of the 75/25 No.1 sample. The morphology details of the single P(VDF-TrFE) electrospinning nanofibers with 75/25 molar rations in this experiment are illustrated in Table 1. In order to ensure the reliability of our experimental results, we prepared two 75/25 P(VDF-TrFE) nanofibers with different diameters and characterize the polarization tunable thermal conductivity respectively.

**Table 1** the diameters, the thermal conductivities and the $\varphi$ results of phonon renormalization of P(VDF-TrFE) nanofibers with different molar ratios.

| Sample | Diameter (nm) | $\kappa_0$ (Wm$^{-1}$K$^{-1}$) | Positive field $\varphi$ | Negative field $\varphi$ |
|---|---|---|---|---|
| 75/25 No. 1 | 178 | 0.23 | 3.35 ± 0.3 | 4.4 ± 0.4 |
| 75/25 No. 2 | 158 | 0.27 | 1.6 ± 0.15 | 1.8 ± 0.2 |
| 70/30 No. 1 [38] | 138 | 1.52 | 3.7 ± 0.5 | 3.7 ± 0.4 |
| 70/30 No. 2 [38] | 511 | 1.10 | 7.9 ± 0.6 | 5.5 ± 0.7 |

The thermal conductivity of 75/25 P(VDF-TrFE) nanofibers was measured as the polarization electric field increased. Here we focused on the mechanism of thermal transport induced by the polarization electric field. As shown in Fig. 2(c), when the polarization electric field reaches ±20 V，the thermal conductivity of the two samples exhibit the different increments. The thermal conductivity of the No.1 nanofiber approaches to 0.23 Wm$^{-1}$K$^{-1}$ under the zero field, which increases by 4.3% for the 20 V and

3.5% for the -20 V. Consistent experimental conditions are provided for the thermal conductivity measurement of No. 2 nanofiber and we found the zero field thermal conductivity is 0.27 Wm$^{-1}$K$^{-1}$. For the No.2 nanofiber, the increments of thermal conductivity approach to 1.9% and 2.0% for the 20 V and -20 V polarization voltage. The differences under positive and negative polarized electric fields are obviously observed, which is believed due to the inherent hysteresis property of ferroelectric materials. In our experiments, the highest electric potential gradient is 20 MV/m, which is much lower than the coercive electric field of P(VDF-TrFE) ~55 MV/m. Similar enhancement of thermal conductivity in P(VDF-TrFE) film has been observed in Deng's work, and the thermal conductivity under 80 MV/m potential gradient presented 53% higher than film[47]. Compared with previous inorganic crystalline Pb(Zr$_{0.3}$Ti$_{0.7}$)O$_3$ film[48], the increment of thermal conductivity is 0.23%/MV/m under polarization voltage, which is similar with the increment (~0.22%/MV/m) in our P(VDF-TrFE) nanofiber. Our P(VDF-TrFE) amorphous nanofibers could obtain such a prominent increment in thermal conductivity under the polarized electric field like crystalline ferroelectrics. We believe this increment in P(VDF-TrFE) nanofibers might influenced by the phonon renormalization which will be discussed later.

In addition, we found that the thermal conductivities of nanofibers with 70/30 molar ratio present much higher values than that of 75/25 molar ratio nanofibers whose value is still in the range of that in bulk polymers, i.e. ~0.1-0.3 Wm$^{-1}$K$^{-1}$. P(VDF-TrFE) with 70/30 molar ration has the higher mole fraction of TrFE and is generally considered to have better ferroelectric properties (see Table 1). It is possible that there is thermal percolation in P(VDF-TrFE) nanofiber with 70/30 molar ratio, which needs to be further verified by more experiments in the future[49].

We now turn to understand the physics behind the electric field dependent thermal conductivity. Phonon renormalization is believed to be responsible for the measured thermal conductivity. According to the phonon renormalization, the molecules of the ferroelectric polymers will deviate from the equilibrium position under the polarization, and the change of bond length and bond energy become the main factors affecting the phonon transport in P(VDF-TrFE) nanostructures. Based on the phonon renormalization model, the thermal conductivity of ferroelectric polymer under polarization electric field can be expressed by the following formula[38]

$$\kappa = \kappa_0\sqrt{1 + \varphi E} \tag{1}$$

where $\kappa_0$ is the thermal conductivity of 75/25 P(VDF-TrFE) nanofibers under zero field. $E$ is the polarization electric field and $\varphi$ is related to the ferroelectricity of nanofiber. The higher $\varphi$ indicates better ferroelectricity.

We compared the $\varphi$ results of previous P(VDF-TrFE) nanofibers with 70/30 molar ratio (show in Table 1). There are two importance observations: (1) it can be concluded that the 75/25 P(VDF-TrFE) with smaller $\varphi$ have poor ferroelectricity, which is consistent with previous results[50, 51]. (2) nanofibers with larger diameter hold the higher $\varphi$ for both positive and negative electric field. The larger diameter of ferroelectric polymer nanofiber contains more polar phase ($\beta$-phase), and the effect of phonon renormalization of molecular chains become more obvious.

Polarization electric field should affect both bond length, bond energy and bond angle. In the phonon renormalization model, the change of band angle is not taken into account, but it is undeniable that the thermal conductivity of the P(VDF-TrFE) nanofiber would be affected by the change of bond angle[52].

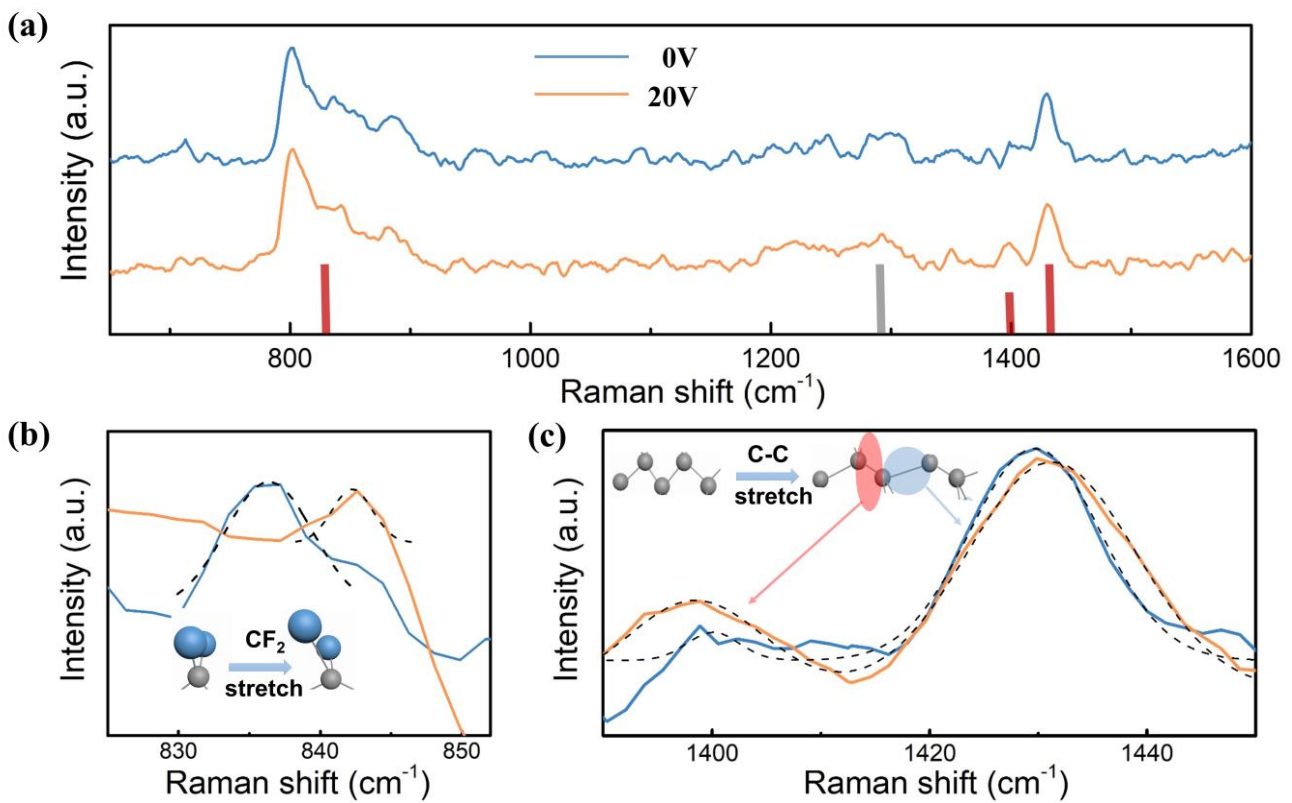

**Figure 3** Raman spectra of the suspended single P(VDF-TrFE) nanofiber with 75/25 molar ratio. The blue lines represent the Raman spectra under 0 V and the orange lines represent that under 20 V. (a) Raman spectra of P(VDF-

TrFE) nanofiber with Raman shift in the range from 650 cm$^{-1}$ to 1600 cm$^{-1}$; (b) Enlarged wavenumbers ranging from 825 cm$^{-1}$ to 852 cm$^{-1}$, among which the peak of CF$_2$ bond is observed. The result illustrates an obvious blue shift under polarization electric field and indicates the bond length increase of CF$_2$ bond; (c) Raman spectra in the range of C-C bond. After applying the polarization electric field, the peak near 1400 cm$^{-1}$ exhibits the red shift that represents the C-C bond length becomes shorter while the peak near 1429 cm$^{-1}$ appears blue shift and represents the C-C bond length becomes longer. The dotted dark lines are peak position fitting.

In order to confirm the prediction of bond length and bond energy in phonon renormalization model, here the Raman spectra was characterized to determine the change of molecular bonds under polarization electric field. Raman spectra is one of the most common methods to characterize molecular bond energy, bond vibration and bond rotation[53, 54]. For fluorine-based ferroelectric polymers (PVDF and P(VDF-TrFE)), non-polar phase (α-phase) and polar phase (β-phase) could be directly resolved from Raman feature bonds according to the previous results[53]. In order to clarify the variation of bond length of P(VDF-TrFE) under the polarization electric field ($E$), Raman responses was study with 20V electric field and 0V electric field, respectively (shown in Figure 3). The shift of Raman peaks can clearly reflect the changes of bond length and bond energy of P(VDF-TrFE) molecular chains. It is no dispute that the electric field only affects the polar phase and thus we should only focus on the change of β-phase in the experimental results. To realize the added polarization field between the two ends of nanofiber, we choose two middle electrodes of MEMS device to apply voltage, as indicated in Fig. 2(a).

**Table 2** Vibrational assignment of P(VDF-TrFE) in the range 650-1600 cm$^{-1}$. The uncertainty of the vibrational frequencies is obtained from the Gauss fit results.

| | β-phase P(VDF-TrFE) | | | |
|---|---|---|---|---|
| Vibrational assignment[54] | theory[32] $E = 0V$ (cm$^{-1}$) | experiment [53] $E = 0V$ (cm$^{-1}$) | this work $E = 0V$ (cm$^{-1}$) | this work $E = 20V$ (cm$^{-1}$) |
| CF$_2$ stretch | 840 | 845 | 836 ± 0.42 | 842 ± 0.80 |
| CH$_2$ scissoring & CF$_2$ stretch | 1281 | 1289 | / | 1292 |
| C-C stretch & CH$_2$ wag | 1409 | 1402 | 1400 ± 0.17 | 1397 ± 0.26 |
| C-C stretch & CH$_2$ scissoring | 1442 | 1431 | 1429 ± 0.30 | 1431 ± 0.12 |

Figure 3 presents the Raman spectra of a single 75/25 P(VDF-TrFE) nanofiber under different polarization electric field (0 V and 20 V). Fig. 3(a) shows the Raman transition of 75/25 P(VDF-TrFE) nanofiber in the range 650-1600 $cm^{-1}$, which is in the range of *β*-phase. Fig. 3(b) is the enlarged Raman spectra in the range from 825 $cm^{-1}$ to 852 $cm^{-1}$, we find an obvious peak with wavenumber at 836 $cm^{-1}$ under 0 V (represent by blue line). 840 $cm^{-1}$ is the same typical characteristic peak of *β*-phase in theory calculation, which represents the $CF_2$ stretch vibration[53, 55]. After adding the polarization electric field ($E$ = 20 V), the 836$cm^{-1}$ peak exhibits a significant blue shift to 842 $cm^{-1}$ (represent by orange line). This blue shift reflects that the bond energy of $CF_2$ bond appears increasing trend. The enhancement of $CF_2$ bond energy is due to the bond length stretched by polarization electric field, which directly increase the bond energy of $CF_2$ bond under stress. The bond length stretched by polarization electric fields could also found in previous work. Viviana and co-workers testify the atomic displacement (C-C and C-N bond) caused by the oriented external electric field in the Indigo molecule[56]. And Vladimir's work also give clear evidence for the change of bond length under electric field in the PVDF system[57].

The change of $CF_2$ bond length is shown in the schematic diagram of Fig. 3(b). Fig. 3(c) presents the change of carbon-carbon molecular bonds. In theory, 1409 $cm^{-1}$ is a characterize peak of C-C stretch and $CH_2$ wag vibration and 1442 $cm^{-1}$ peak presents the C-C stretch and $CH_2$ scissoring vibration. Neither $CH_2$ wag nor $CH_2$ scissoring will change the bond length and thus these vibrations don't produce peak shift. In our experiments we find the corresponding peaks at 1400 $cm^{-1}$ and 1429 $cm^{-1}$ under 0 V field. After adding electric field, the 1400 $cm^{-1}$ peak appears red shift to 1397 $cm^{-1}$ that exhibits the decreased bond energy only in C-C bond and the 1429 $cm^{-1}$ peak shows blue shift to 1431 $cm^{-1}$ that present the increased C-C bond energy. The change of wavenumbers of 1400 $cm^{-1}$ peak and 1429 $cm^{-1}$ peak expound that the carbon atom will deviates from the equilibrium position under the effect of polarization electric field corresponding to the phonon renormalization model, and the bond length of the adjacent C-C bond will increase and decrease (shows in Fig. 3(c)). In addition, we also found a Raman peak around 1292 $cm^{-1}$ that can characterize the change of $CH_2$ bond angle after adding electric field.

The peak shift of C-C bond confirms that the phonon renormalization appears in P(VDF-TrFE)

nanofiber. The applied polarization electric field leads to a red shift of the peak near 1400 cm$^{-1}$ and a blue shift near 1429 cm$^{-1}$ of C-C bond, which has been intuitively shown in the inset schematic diagram of Fig. 3(c). From the schematic diagram, we find the length of adjacent C-C bonds in the P(VDF-TrFE) backbone cannot be longer or shorter at the same time under the strength of the polarization electric field. The bond length of C-C bond presents shorter in red circle and becomes longer in blue circle. This difference of C-C length is due to the carbon atoms will be displaced from the original equilibrium positions in the process of phonon renormalization (caused by applied polarization electric field) and the adjacent carbon atoms will move in the opposite direction. The harmonic approximation of the potential at the new equilibrium position will be different from the original one due to the anharmonicity. Therefore, the lattice constant and bond energy are different from that in unpolarized molecular bonds.

In conclusion, the polarization electric field could induce the changes of bond energy, bond length and bond angle of the ferroelectric polymers. The change of bond length and bond energy is the key point of the increased thermal conductivities of the P(VDF-TrFE) nanofibers, evidenced from the Raman which directly prove the polarization induced phonon renormalization of fluorine-based polymers. In addition, the 75/25 P(VDF-TrFE) nanofiber contains inferior ferroelectricity comparing to 70/30 P(VDF-TrFE) nanofiber and the large diameter nanofiber will have more polar phases and better ferroelectricity.

**METHODS**

**The preparation of suspended MEMS device**

The supported MEMS device has been processed by previous micro-nano procedures, including ultraviolet lithography, metal deposition (Pt electrodes deposited on the SiNx beam), lift-off and deep reactive ion etching. After Potassium hydroxide (KOH solution with 30% volume ratio) wet etching, the suspended MEMS devices are fabricated before electrospining process. KOH solution is mainly used for etching silicon oxide substrate which is not be covered by Pt/SiNx electrodes. In order to prevent the suspended Pt/SiNx electrodes from being broken by the tension of the solution (KOH solution, deionized water, etc.), in our experiments we used a critical dryer (Tousimis Samdri-PVT-3D) to dry the MEMS device and protect the suspended electrodes. After KOH wet etching, the whole suspended MEMS devices are annealed in $H_2$/Ar atmosphere at around 250°C for 2-3 hours to remove the water molecules and polymer residues on the surface of the device. Electrospinning process is carried out immediately after $H_2$/Ar atmosphere annealing to prevent the accumulation of water molecules on the electrode from affecting the thermal contact between the P(VDF-TrFE) nanofiber and the Pt/SiNx electrodes.

**Electrospining**

The electrospinning solvent, mixture of P(VDF-TrFE) powder and Dimethylformamide (DMF) solution, was prepared with 12 wt% to 18 wt%. Usually, P(VDF-TrFE) solution should be stirred at room temperature for more than 12 hours to ensure the complete dissolution of P(VDF-TrFE) powder. Considering we need to apply a large enough potential gradient to both ends of P(VDF-TrFE) nanofiber, here we chose the suspended MEMS devices with a gap no more than 3μm between the two thermometers to fabricate the suspended P(VDF-TrFE) nanofibers. We fixed the syringe speed (0.5 mL/h) and the distance (20 cm) which between needle tip and the MEMS device during the electrospinning process, and we found the diameter of P(VDF-TrFE) nanofibers is related to the high voltage (9-13 kV) and solution concentration. To fabricate a suspended single P(VDF-TrFE) nanofiber across the two thermometers and the electrodes which used to add the polarization electric field, the electrospinning time is controlled within 10 seconds.

**Thermal bridge method**

The thermal conductivity of a suspended single P(VDF-TrFE) nanofiber is characterized by thermal bridge method. In order to reduce the effect of thermal convection, the whole suspended MEMS device with a single P(VDF-TrFE) nanofiber was placed into a cryostat with high vacuum on the order of $1\times10^{-4}$ Pa. We waited for at least 2 hours at each temperature point to ensure the stability of the ambient temperature. To provide heating power, a 10 μA DC current (provided by Keithley 6221) is added to the one thermometer severed as heater to produce the total joule heat. After reaching thermal equilibrium, the same AC currents (~1 μA provide by Keithley 6221) are used to observe the change of resistance of the two thermometers which act as heater mentioned above and sensor. The change of voltage of the heater and sensor are measured by Lock-in amplifier (SRS 830). The change of resistance of thermometers could characterize the temperature rise ($\Delta T_h$ and $\Delta T_s$) of the both ends of the P(VDF-TrFE) nanofiber. Here the thermal conductance of the P(VDF-TrFE) nanofiber

($G_{\text{P(VDF-TrFE)}}$) could be calculated by the thermal equilibrium formula:

$$G_{\text{b}} = \frac{Q_{\text{tot}}}{\Delta T_{\text{h}} + \Delta T_{\text{s}}}$$

$$G_{\text{P(VDF-TrFE)}} = \frac{G_{\text{b}} \Delta T_{\text{s}}}{\Delta T_{\text{h}} - \Delta T_{\text{s}}}$$

where $Q_{\text{tot}}$ is total joule heat add on the heater, $G_{\text{b}}$ is the thermal conductance of the Pt/SiNx beams, $\Delta T_{\text{h}}$ and $\Delta T_{\text{s}}$ act as the temperature rise of the heater and sensor respectively. The thermal conductivity a single P(VDF-TrFE) nanofiber could be obtained by:

$$\kappa = G_{\text{P(VDF-TrFE)}} \frac{L}{A}$$

where $\kappa$ is the thermal conductivity of the P(VDF-TrFE) nanofiber, $L$ and $A$ is the length and cross section area of the P(VDF-TrFE) nanofiber.

## ACKNOWLEGEMENT

The work was supported by the National Natural Science Foundation of China (No. 12004242 & 11890703 & 12174286 & 11935010 & 51876111), by the Key-Area Research and Development Program of Guangdong Province (No. 2020B010190004), by Shanghai Rising-Star Program (No. 21QA1403300) and by Shanghai Local Capacity Building Program (No. 22010500700).